\documentstyle[sprocl,epsfig,psfrag,epsfig,axodraw,amsmath,feynarts]{article}

\input paperdef

\begin{document}
\begin{flushright}
\parbox{10cm}{
\vspace{-3cm}
\begin{flushright}
IPPP/04/40\\
DCPT/04/80\\
CERN--PH--TH/2004--129
\end{flushright}
}
\end{flushright}
\vspace{-2em}
\title{The anomalous  magnetic
  moment of the muon in the MSSM --- recent developments}

\author{Sven Heinemeyer$^a$, Dominik St\"ockinger$^b$, and Georg
  Weiglein$^b$}

\address{
$^a$ CERN, TH Division, Dept. of Physics, 1211 Geneva 23,
  Switzerland\\
$^b$ Institute for Particle Physics 
Phenomenology, University of Durham, UK
}

\maketitle
\abstracts{
We present recent results of two interesting classes of supersymmetric
two-loop contributions to $(g-2)_\mu$. Two-loop diagrams
involving either a closed sfermion loop or a closed
chargino/neutralino loop can amount to $5\times10^{-10}$, which
is almost one standard deviation of the current experimental
uncertainty. We discuss the dependence of these two classes on
the unknown supersymmetric parameters and their impact on the
supersymmetric prediction of $(g-2)_\mu$.
}

\section{Introduction}

After continuous improvement in the experimental
\cite{g-2exp} and Standard Model-theoretical
\cite{DEHZ,g-2HMNT,Jegerlehner,Yndurain,LBL,LBLnew} determination of
the anomalous magnetic moment $\amu=(g-2)_\mu/2$ of the muon, there
remains a tantalizing discrepancy
\footnote{Here we use the evaluations from \cite{g-2HMNT,LBLnew} for
  the hadronic contributions. Other $e^+e^-$ data driven evaluations
  result in similar deviations of $2-3\si$.  
Recent analyses concerning $\tau$ data indicate that uncertainties due to
isospin breaking effects may have been underestimated
earlier~\cite{Jegerlehner}. We thank F.\ Jegerlehner for discussions
  on this point.}
\BEA
\label{deviationfinal}
\amuexp-\amu^{\rm theo,SM} & = &
(24.5\pm9)\times10^{-10}
\EEA
between the experimental value and the Standard Model prediction.

It is an interesting question whether the observed deviation
(\ref{deviationfinal}) is due to supersymmetric effects. The
supersymmetric one-loop contribution
is approximately given by \cite{Moroi}
\BE
\amu^{\SU,{\rm 1L}} \approx 13 \times 10^{-10} 
  \frac{\tb\, {\rm  sign}(\mu)}{\left(\msusy/100\gev\right)^2} ,
\label{susy1loop}
\EE
if all supersymmetric particles (the relevant ones are the smuon,
sneutralino, chargino and neutralino) have a common mass
$\msusy$. 

This formula shows that supersymmetric
effects can easily account for a 
$(20\ldots30)\times10^{-10}$ deviation, if $\mu$ is positive and
$\msusy$ lies roughly between 100 GeV (for small $\tb$) and
600 GeV (for large $\tb$).
On the other hand, the precision of the
measurement places strong bounds on the supersymmetric parameter
space. 


Here we review the results of \citeres{g-2FSf,g-2ChaNeu} for
the Minimal Supersymmetric Standard Model (MSSM) two-loop
contributions of
\begin{list}{---}{
\addtolength{\itemsep}{-1ex}
\addtolength{\topsep}{-1ex}
\addtolength{\leftmargin}{-1.5ex}}
\item two-loop diagrams involving a closed subloop of sfermions (stops,
  sbottoms, staus, and tau-sneutrinos)
\item two-loop diagrams involving a closed subloop of charginos and/or
  neutralinos
\end{list}
These contributions constitute the class of
two-loop contributions to $\amu$, where a supersymmetric loop is
inserted into a SM (or more precisely a two-Higgs-doublet model)
one-loop diagram.  



These diagrams are particularly interesting
since they can depend on other parameters than the supersymmetric
one-loop diagrams and can therefore change the qualitative behaviour
of the supersymmetric contribution to $\amu$. In particular, they could
even be large if the one-loop contribution is suppressed, e.g.\ due to
heavy smuons and sneutrinos.

Calculational details and remarks to the regularization and the
$\gamma_5$ problem can be found in
\citeres{g-2FSf,g-2ChaNeu,procLL2004}. Essentially we evaluate the
two-loop and corresponding counterterm diagrams using standard large
mass expansion and integral reduction techniques.
A major difficulty stems from the large number of
different mass scales and the involved structure of the MSSM Feynman
rules.

\section{Parameter dependence and discussion}

The results for the supersymmetric contributions to $\amu$ are 
functions of all MSSM parameters. The values of the MSSM parameters
are unknown, but the parameter space is strongly restricted by several
experimental constraints. It turns out 
the parameter dependence and the corresponding phenomenological
discussion shows important differences between the sfermion
and the chargino/neutralino loop contributions.

\begin{list}{---}{
\addtolength{\itemsep}{-1ex}
\addtolength{\topsep}{-1ex}
\addtolength{\leftmargin}{-1.5ex}}
\item
The sfermion loop contributions depend on the Higgs sector parameters
$\mu$ and $\tan\beta$ and the sfermion mass parameters in a rather
complicated way. It also turns out that experimental
constraints on the MSSM parameter space significantly restrict the
possible sfermion loop contributions \cite{g-2FSf}. 
\item
In contrast, the chargino/neutralino loop contributions depend on
$\mu$, $\tan\beta$ and the gaugino mass parameter $M_2$ in a quite
straightforward way, and experimental constraints on the parameter
space have not much impact \cite{g-2ChaNeu}. 
\end{list}

\subsection{Sfermion contributions}

\begin{figure}
\BC
\epsfig{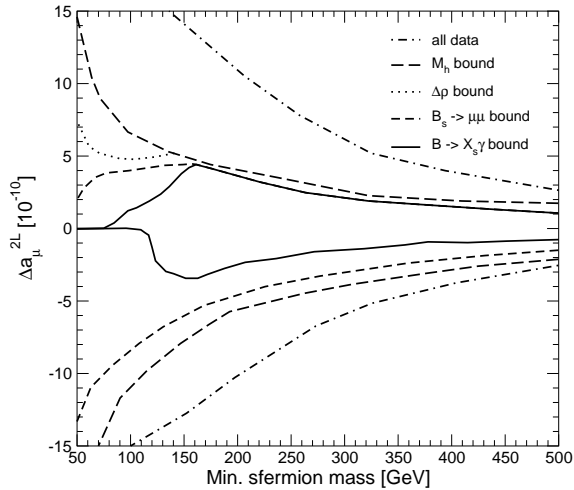}
\vspace{-1em}
\caption{%
Maximum contributions of the two-loop diagrams with a closed sfermion loop 
to $\amu$ as a function of the lightest sfermion mass. No constraints on the
MSSM parameters are taken into account for the outermost curve. Going
to the inner curves additional constraints (see text) have
been applied.
}
\label{fig:gammaWscan}
\EC
\vspace{-2em}
\end{figure}

In \reffi{fig:gammaWscan} we show the full results for the sfermion
contributions as functions of the lightest sfermion mass for universal
sfermion mass parameters. They are obtained from a scan over the
supersymmetric parameter space and display clearly the impact of
taking into account experimental constraints on the parameter space. 

The outer
lines show the maximum possible results for $\tan\beta=50$ if all MSSM
mass parameters are varied universally up to $3\tev$ (for the $\cp$-odd
Higgs-boson mass we use $M_A>150\gev$) ignoring all
experimental constraints on the parameter space. The next lines show the
maximum possible results if only parameter points are used that are in
agreement with the experimental limit on
$M_h$ \footnote{For a full list of references on the experimental
  constraints see \citere{g-2FSf}.}. As indicated above we find that
the maximum results are drastically reduced. For a 
lightest sfermion mass of $100\gev$, the results are reduced from more
than $15\times10^{-10}$ to about $5\times10^{-10}$. The inner lines
correspond to taking into account more experimental constraints on
$\De\rho$, $\br(B_s\to\mu^+\mu^-)$ and 
  $\br(B \to X_s\ga)$. They reduce the maximum
contributions further.

As discussed in detail in \citeres{g-2FSf,procLL2004}, the restriction
of universal sfermion mass parameters used in \reffi{fig:gammaWscan} is
indirectly responsible for the significant impact of the
$\Mh$-bound. If the ratio between sbottom and stop masses is very
large and one stop mass remains light, larger contributions to $\amu$
become possible without violating the experimental bounds.


\subsection{Chargino/neutralino contributions}

The chargino/neutralino two-loop contributions have a more straightforward
parameter dependence. They depend on $\tan\beta$ and the mass
parameters for the Higgsinos, $\mu$, the gauginos, $M_2$, and the
$\cp$-odd Higgs boson, $M_A$. For the simple case that all these mass
parameters are equal to a common mass scale $\msusy$, we obtain the
approximation
\BE
\amu^{\chi,\rm 2L} = 11 \times 10^{-10} 
  \frac{(\tb/50)\, {\rm  sign}(\mu)}{\left(\msusy/100\gev\right)^2} .
\label{chaneuapprox}
\EE
If all the masses are even equal to the smuon and sneutrino masses,
this formula can be immediately compared to the one-loop contributions
(\ref{susy1loop}). In this case the chargino/neutralino two-loop
contributions amount to about 2\%\ of the one-loop contributions.

\begin{figure}[htb]
\vspace{-1em}
\unitlength=21.0bp%
\begin{picture}(10,8.5)\epsfxsize=5.5cm
 \put(000,000.5){\epsfxsize=5.5cm\epsfbox{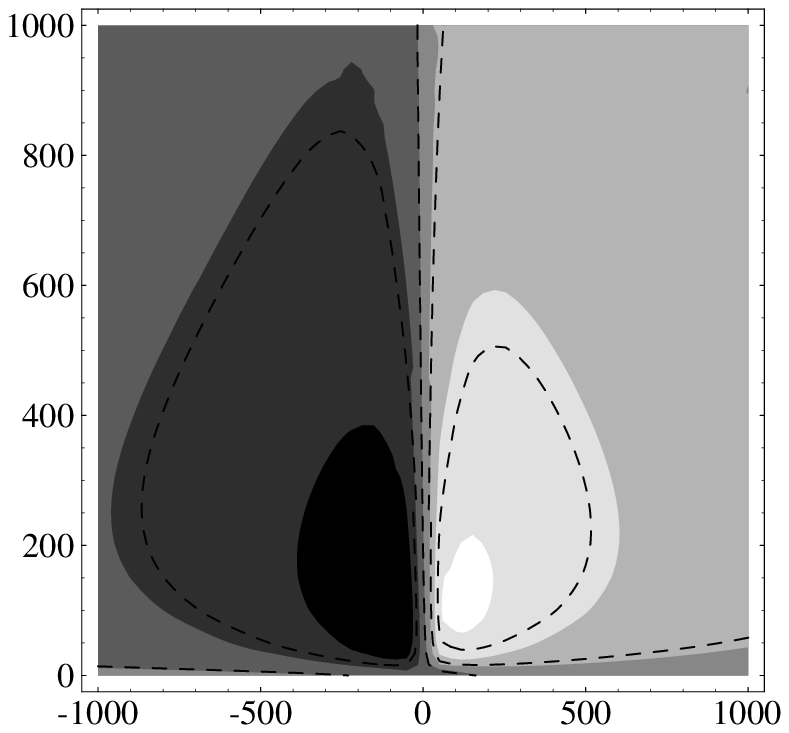}}
 \put(8,0.5){\epsfxsize=5.5cm\epsfbox{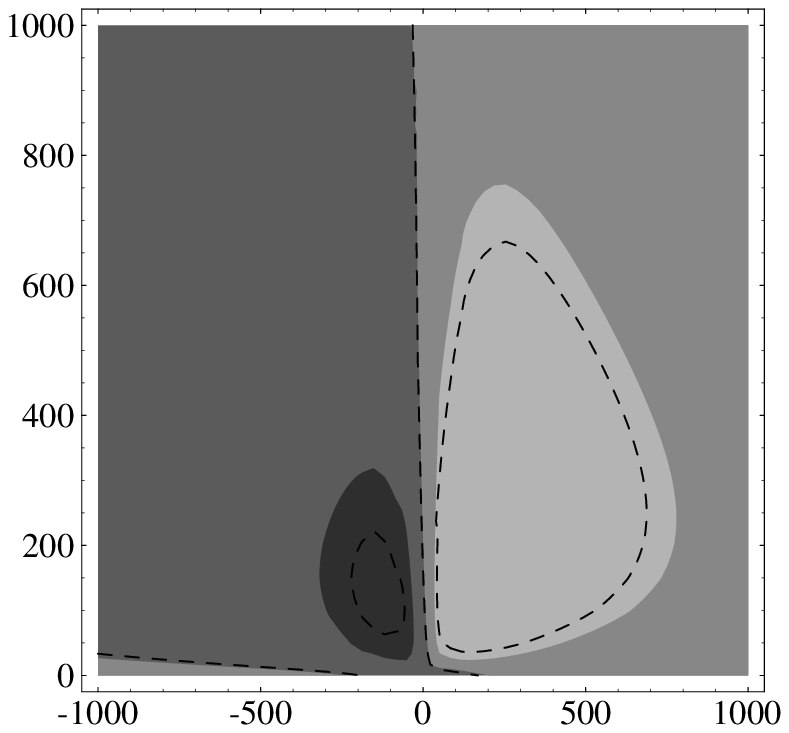}}
\put(0.25,7.7){\footnotesize $M_2$ [GeV]\qquad\quad$\tan\beta=50$}
\put(8.25,7.7){\footnotesize $M_2$ [GeV]\qquad\quad$\tan\beta=25$}
\put(7.25, 0.25){\footnotesize $\mu$ [GeV]}
\put(15.25,0.25){\footnotesize $\mu$ [GeV]}
\put(4.6,2.5){\footnotesize $<1\si$}
\put(5.,5.5){\footnotesize $1-2\si$}
\put(1.5,6.5){\footnotesize $3-4\si$}
\put(12.6,2.7){\footnotesize $1-2\si$}
\put(13,6.5){\footnotesize $2-3\si$}
\put(9.5,6.5){\footnotesize $3-4\si$}
\end{picture}
\vspace{-1em}
\caption{Contour plots of $\amu^{\SU,\rm 1L}+\amu^{\chi,\rm
  2L}$ (fully drawn areas) and $\amu^{\SU,\rm 1L}$ (dashed contours)
  in the $\mu$--$M_2$-plane. In the left plot we choose
  $\tan\beta=50$, in the right plot $\tan\beta=25$. The borders of the
  regions and the 
  contours correspond to $1\si$, $2\si$, \ldots deviation from the
  observed value according to eq.\ (\ref{deviationfinal}).}
\vspace{-2em}
\label{fig:chaneuplot}
\end{figure}

If the smuon and sneutrino masses are heavier than the chargino and
neutralino masses, the one-loop contributions are suppressed and the
two-loop contributions can have a larger
impact. Fig.~\ref{fig:chaneuplot} shows the sum $\amu^{\SU,\rm
  1L}+\amu^{\chi,\rm 
  2L}$ in comparison to the one-loop result $\amu^{\SU,\rm 1L}$ alone
as a contour plot in the $\mu$--$M_2$-plane. The smuon and sneutrino
masses are fixed to $1\tev$ and $\tb=50,25$, $M_A=200$. We find that in
this case the two-loop corrections from the chargino/neutralino loop
diagrams can modify the $1\si$, $2\si$, \ldots contours significantly.

\section{Outlook}

Supersymmetric contributions to $\amu$ could easily account for the
observed $(20\ldots30)\times10^{-10}$ deviation between SM theory and
experiment. Conversely, the precision of the experiment places
stringent bounds on the MSSM parameter space.

The two-loop contributions presented here can substantially modify the
supersymmetric one-loop contribution, and their knowledge reduces the
theoretical uncertainty of the supersymmetric prediction for
$\amu$. Apart from the magnitude of these contributions (of order
$0.5\ldots1\si$), it is
interesting how significantly the experimental constraints on the
supersymmetric parameter space influence the possible results.

In \citeres{g-2FSf,g-2ChaNeu} also the SM/two-Higgs-doublet model-like
contributions of two-loop diagrams with fermion loops and with purely
bosonic loops have been computed. The difference of the diagrams in
the MSSM and the SM is smaller than $1\times10^{-10}$. The remaining
task is to complete the full two-loop calculation of $\amu$ in the
MSSM. The missing diagrams are the two-loop corrections to the
supersymmetric one-loop diagrams with smuon or sneutrino exchange. In
order to calculate them, the full one-loop renormalization of the MSSM
will be needed.

\end{document}